\documentclass{article}

\usepackage{amsfonts}
\usepackage[intlimits]{amsmath}
\usepackage{cite}
\usepackage{epsfig}

\def\bea{\begin{eqnarray}}
\def\eea{\end{eqnarray}}
\def\nn{\nonumber}

\def\lmatrix{\left(\begin{array}}
\def\rmatrix{\end{array}\right)}
\def\theta{\vartheta}

\begin{document}

\vspace{1cm}

\begin{center}

{\Large\bf An ideal toy model for confining,}


\vspace{0.5cm}


{\Large\bf walking and conformal gauge theories: }

\vspace{0.5cm}

{\Large\bf the $O(3)$ sigma model with $\theta$-term }

\vspace{2cm}

D\'aniel N\'ogr\'adi

\vspace{1cm}

{\em E\"otv\"os University, Department for Theoretical Physics}

\vspace{0.2cm}

{\em P\'azm\'any P\'eter s\'et\'any 1/a, Budapest 1117, Hungary}

\vspace{1cm}

{\tt nogradi@bodri.elte.hu}

\end{center}

\vspace{2cm}

\begin{abstract}

A toy model is proposed for four dimensional non-abelian gauge theories coupled to a large number of fermionic degrees of freedom. As the number
of flavors is varied the gauge theory may be confining, walking or conformal. The toy model mimicking this feature is the two dimensional $O(3)$ 
sigma model with a $\theta$-term. For all $\theta$ the model is asymptotically free. For small $\theta$ the
model is confining in the infra red, for $\theta = \pi$ the model has a non-trivial infra red fixed point and
consequently for $\theta$ slightly below $\pi$ the coupling walks. The first step in investigating the notoriously
difficult systematic effects of the gauge theory in the toy model is to establish non-perturbatively that the $\theta$ parameter
is actually a relevant coupling. This is done by showing that there exist quantities that are entirely given by the total
topological charge and are well defined in the continuum limit and are non-zero, despite the fact that the topological
susceptibility is divergent. More precisely it is established that the differences of connected correlation functions
of the topological charge (the cumulants) are finite and non-zero and consequently there is only a single divergent
parameter in $Z(\theta)$ but otherwise it is finite. This divergent constant can be removed by an appropriate counter term
rendering the theory completely finite even at $\theta > 0$.


\end{abstract}

\newpage

\section{Introduction}
\label{introduction}

Lattice simulations of technicolor inspired models are plagued by known systematic uncertainties \cite{DelDebbio:2011kp,
Catterall:2011ce, Bursa:2011ru, Sint:2011gv, Fodor:2009wk}. Although the models under consideration are
QCD-like in that they are four dimensional non-abelian gauge theories coupled to dynamical fermions the systematic effects of the
interesting models (those that are either conformal or walking) are much more
difficult to control than in actual QCD. As a result currently there are disagreements between various approaches,
discretizations, etc, and universality is not immediately evident \cite{Appelquist:2009ty, Jin:2009mc, Deuzeman:2009mh,
Fodor:2011tu, Hasenfratz:2011xn}. Clearly the general expectation is that once all systematic
effects are controlled and taken into account the results from different approaches and regularizations will agree as they should. 

In this paper a toy model is proposed which mimics many of the features of non-abelian gauge theories in the hope that systematic
effects can be fully explored. Hopefully these will help controlling the corresponding effects in the much more complicated
gauge theories. The proposed model is the
two dimensional $O(3)$ non-linear sigma model with a $\theta$ term.
At $\theta = 0$ the model served as a toy model of QCD for a long time
since it is asymptotically free, features instantons, confinement and dimensional transmutation \cite{Novikov:1984ac}. 
It is exactly solvable \cite{Zamolodchikov:1978xm} even at finite volume \cite{Balog:2003yr, Balog:2005yz, Balog:2009ze}. 
Since the topological term is invisible in perturbation theory the model is asymptotically free for
arbitrary $\theta$. The dynamics in the infra red is however expected to be very sensitive to $\theta$.

At $\theta = \pi$ the model is conjectured \cite{Haldane:1982rj, Haldane:1983ru}
to have a non-trivial infra red fixed point governed by the $SU(2)$ WZNW model at level $k=1$ and, if the conjecture holds, 
is also exactly solvable. Some numerical evidence in support of the conjecture has been presented in \cite{hep-lat/9505019} and a
recent very detailed study confirming it in \cite{Bogli:2011aa}. The
infra red fixed point implies a zero of the $\beta$-function. This situation is analogous to gauge theories in the conformal
window.

For $0 < \theta < \pi$ exact solvability is lost but based on continuity one expects that for $\theta$ not much below $\pi$ the
$\beta$-function develops a near zero and the renormalized coupling will walk. This arrangement is analogous to gauge theories just
below the conformal window. Hence dialing $\theta$ corresponds to dialing the number of flavors $N_f$ in the gauge theory. 

In all three scenarios (confining, walking, conformal) one may 
also introduce an external magnetic field to mimic the effect of a finite quark mass.

Before exploring the analogies further and investigating the origins of the severe systematic effects
the first task is to establish non-perturbatively that the $\theta$-term is actually a relevant operator and also what the
singularity structure of the theory is for $\theta > 0$.
This is not immediately obvious largely because of the unusual scaling
properties of the topological susceptibility and a class of similar observables.

It is well known that small size instantons render the topological susceptibility $\chi =
\langle Q^2 \rangle / V$ ill defined in the semi-classical approximation \cite{Luscher:1981tq}. Going beyond the semi-classical
approximation fully non-perturbative lattice studies have shown that regardless how one improves the details of the lattice implementation
a logarithmically divergent susceptibility is obtained at finite physical volume in the continuum limit. Moreover, all even
moments of the total topological charge distribution $\langle Q^{2m} \rangle / V$ have the same property.

However, the model at $\theta = 0$ is exactly solvable and both the exact solution and the continuum limit of lattice simulations agree that correlators of
the topological charge density, e.g. $\langle q(x) q(0) \rangle$ are finite. The above two observations, namely that certain
statistical properties of the total charge distribution $P(Q)$ are ill defined while at the same time correlators of $q(x)$ are
finite, might make one wonder whether the total charge operator $Q$ is an irrelevant operator while $q(x)$ is not. If so, the only consistent
continuum value of $\langle Q^{2m} \rangle$ would be zero and the apparent divergences in the lattice calculations would be
regarded as artifacts. This scenario would imply that the theory defined on the lattice at non-zero $\theta$ leads to an
identical continuum theory as the one defined at $\theta = 0$. Equivalently,
the total charge operator inserted into any correlation function would be zero in the continuum
theory $\langle Q \ldots \rangle = 0$, while correlation functions of the type $\langle q(x) \ldots \rangle$ are finite. This
scenario would of course invalidate Haldane's conjecture about the equivalence of the $\theta = \pi$ theory with a non-trivial
interacting conformal field theory.

In this work it is shown that there exist quantities built out of the total topological charge operator $Q$ which have well
defined continuum limits and are non-zero. These observables are differences of connected correlation functions of the topological
charge, in other words the cumulants. Each term
is logarithmically divergent but the divergence cancels in the difference and moreover they scale correctly in the continuum limit
to non-zero values. Showing correct scaling towards the continuum limit in itself would not be sufficient to prove that the
$\theta$-term is a relevant operator because the continuum limit value could be zero. Since all cumulant differences are finite
there is only a single UV-divergent parameter in the partition function $Z(\theta)$ but otherwise it is finite.

While preparing this manuscript the preprint \cite{Bogli:2011aa} appeared also with the conclusion that $\theta$ is a relevant
coupling. The method was different though, in \cite{Bogli:2011aa} it was shown to high precision that a well defined observable is different
in the continuum limit for three different values of $\theta$ implying that $\theta$ can not be irrelevant. In the current work all
simulations are carried out at $\theta = 0$ and the same conclusion is reached by showing that certain combinations of the
topological charge operator are non-zero in the continuum.

\section{$O(3)$ sigma model with a $\theta$-term}
\label{topological}

The model in Euclidean continuum notation is defined by the action
\bea
S = \frac{1}{2g_0^2} \int d^2x \partial_\mu s_a \partial_\mu s_a
\eea
for the unit 3-vectors $s$, $s_1^2 + s_2^2 + s_3^2 = 1$, where $g_0$ is the bare coupling. 
Only a torus geometry will be considered corresponding to a box of finite
linear size $L$ which will be regularized by a symmetric lattice.

The corresponding partition function, free energy per unit volume and topological charge distribution of the model at non-zero $\theta$ and volume $V$ is given by
\bea
Z(\theta) = \langle e^{i\theta Q} \rangle = e^{-Vf(\theta)} = \sum_Q P(Q) e^{i\theta Q}\;,
\eea
with the normalization $Z(0)=\sum_Q P(Q) = 1$. Since physics is periodic with period $2\pi$ in $\theta$ and $\theta\to-\theta$ is
a symmetry the free energy per unit volume can be Fourier expanded
\bea
f(\theta) = \sum_{n=1}^\infty \left( 1 - \cos( n\theta) \right) f_n\;.
\eea

It has been pointed out in \cite{seiberg} that in the semi-classical or dilute gas approximation all $f_n$ coefficients vanish
except for $f_1$ which is UV divergent due to instantons of size $a \ll \rho \ll \xi$ where $a$ is the lattice cut-off and $\xi$
is the physical correlation length.
The remaining coefficients come from interactions between instantons.
Semi-classical arguments also suggest that for instantons causing the UV divergence in $f_1$
the ratio between their size and their average separation goes to zero in the continuum limit. This would imply that the interactions
responsible for the $f_{n>1}$ coefficients are small in the continuum limit hence will not cause them to diverge.

To summarize, the semi-classical approximation accounts for a UV divergent $f_1$ and finite $f_{n>1}$ coefficients. A suitable way
of addressing whether this statement is true beyond the semi-classical approximation is to consider observables that can be
expressed by the $f_{n>1}$ coefficients only and calculating them fully non-perturbatively. The simplest choice is to take the
connected correlation functions of the topological charge,
\bea
\chi_{2m} = (-1)^{m+1} \left. \frac{d^{2m}f}{d\theta^{2m}}\right|_{\theta=0}
\eea
and consider their differences,
\bea
\Delta \chi_{2m} = \chi_{2m} - \chi_{2m+2} = \sum_{n=2}^\infty f_n n^{2m} ( 1 - n^2 )
\eea
from which $f_1$ drops out. The first few such correlation functions are 
\bea
\chi_2 &=& \frac{\langle Q^2 \rangle}{V}  \nn \\
\chi_4 &=& \frac{\langle Q^4 \rangle - 3 \langle Q^2 \rangle^2}{V}\\
\chi_6 &=& \frac{\langle Q^6 \rangle - 15 \langle Q^4 \rangle \langle Q^2 \rangle + 30 \langle Q^2 \rangle^3}{V}\;.\nn
\eea
All of these are expected to diverge in the continuum limit but their differences are expected to be finite. Some numerical evidence
has been presented in \cite{seiberg} in favor of correct scaling behavior for $\Delta\chi_2$ but whether the continuum value is zero
or non-zero has not been discussed.

In the following it will be shown to high precision that the expectations from the semi-classical analysis indeed hold
non-perturbatively and all moments $\langle Q^{2m} \rangle$ and all cumulants $\chi_{2m}$ are logarithmically divergent but the
differences $\Delta \chi_{2m}$ are finite. This implies that there is a single ill-defined constant in $f(\theta)$ namely $f_1$
but otherwise it is finite. The constant $f_1$ can be removed by an appropriate renormalization condition leading to a finite and
universal free energy and partition function for arbitrary $\theta$.

\section{Numerical simulation}
\label{results}

It is convenient to take the continuum limit on a symmetric periodic lattice $L^2$ of fixed physical volume. Physical length and mass is
defined by the second moment correlation length $\xi_2$ \cite{Caracciolo:1992nh},
\bea
\frac{1}{\xi_2(L)^2} = \left(\frac{\sin\frac{\pi a}{L}}{\frac{\pi a}{L}}\right)^2 \left( 2 \frac{M_0}{M_2} -
\frac{4\pi^2}{L^2} \right)
\eea
where
\bea
M_{2n} = \left(\frac{L}{2\pi}\right)^{2n} \sum_t \left( 2 \sin\frac{\pi t}{L} \right)^{2n} C(t)
\eea
is given in terms of the zero spatial momentum projection of the 
2-point correlation function $C(t) = \sum_x \langle s_a(t,x) s_a(0,0) \rangle$ of the field $s$. Let us introduce $m(L) = 1 / \xi_2(L)$.
Note that in this notation $m(L)$ is not the mass gap in finite
volume but rather is simply
defined as the inverse of $\xi_2$ (which for $L\to\infty$ agrees with the mass gap but not for finite
$L$). The physical volume is fixed to $m(L)L = 4$. 
A novel \cite{Patrascioiu:1992na, Hasenbusch:1995hu, arXiv:1009.2146} topological lattice action is
used for the simulations,
\bea
S = \sum_{\langle i, j \rangle} S(s_i, s_j)
\eea
where the sum is over all neighboring sites and 
\bea
S(s_i,s_j) = \left\{\begin{array}{l} 0 {\rm\quad if\quad} s_i \cdot s_j > \cos \delta \\ \infty  {\rm\quad otherwise} \end{array}\right.
\eea
In other words the action is zero for two neighboring vectors if their relative angle is smaller than $\delta$ and infinite
otherwise. The continuum limit is taken by tuning the bare coupling $\delta$ towards zero. This action is topological because small
perturbations of the field $s$ do not change the action nevertheless it has been shown that it is in the right universality
class \cite{arXiv:1009.2146}.

If $\delta < \pi/2$ powerful improvements exist for the measurement of the topological charge distribution \cite{hep-lat/9505019}
based on a generalization of the usual cluster algorithms \cite{Niedermayer:1988kv, Wolff:1988uh}. 
The topological charge operator from \cite{Berg:1981er} is used assigning an integer charge to each configuration even at finite lattice
spacing. 

The continuum extrapolation of the cumulant differences will be done
through 12 lattice spacings using the parameter values from \cite{arXiv:1009.2146} listed in table \ref{tab:parameters}.
The measured correlation lengths and topological susceptibilities are in agreement with those in \cite{arXiv:1009.2146}. In the
present work $O(10^8)$ configurations were generated at each volume and every 10$^{th}$ was measured for the topological charge
distribution and correlation length. The large number of configurations was necessary because there are huge cancellations
between the various terms in the difference of cumulants, especially for $\Delta\chi_4$. The third difference, $\Delta\chi_6$, was
already impossible to obtain with the current statistics.

\begin{table}
\begin{center}
\begin{tabular}{|c|c|c|c|c|c|c|c|}
\hline
$L/a$ & $\delta/\pi$ & $m(L)L$ & $ L^2\chi_2$ & $L^2\chi_4$ & $L^2\chi_6$ & $L^2\Delta\chi_2$ & $L^2\Delta\chi_4$ \\
\hline
\hline
60  & 0.48490 &   4.0017(14)    & 1.2957(2)  & 0.8812(8) & -0.019(5) & 0.4145(8) & 1.069(5) \\
80  & 0.47260 &   4.0032(19)    & 1.4651(2)  & 1.0292(8) & -0.011(6) & 0.4359(7) & 1.143(6) \\
100 & 0.46370 &   4.0007(19)    & 1.6018(3)  & 1.1512(9) & -0.035(8) & 0.4507(9) & 1.186(7) \\
120 & 0.45680 &   3.9939(20)    & 1.7155(3)  & 1.257(1)  &  0.033(9) & 0.459(1)  & 1.224(8) \\
160 & 0.44680 &   4.0011(14)    & 1.9214(4)  & 1.444(1)  &  0.16(1)  & 0.477(1)  & 1.28(1)  \\
200 & 0.43950 &   4.0015(17)    & 2.0836(3)  & 1.596(1)  &  0.24(1)  & 0.488(1)  & 1.35(1)  \\
240 & 0.43385 &   3.9998(14)    & 2.2208(3)  & 1.729(1)  &  0.40(1)  & 0.492(1)  & 1.33(1)  \\
320 & 0.42545 &   4.0010(17)    & 2.4476(4)  & 1.946(1)  &  0.57(1)  & 0.502(1)  & 1.37(1)  \\
400 & 0.41930 &   3.9983(14)    & 2.6259(4)  & 2.118(2)  &  0.71(2)  & 0.508(2)  & 1.41(2)  \\
480 & 0.41455 &   4.0014(19)    & 2.7845(4)  & 2.274(2)  &  0.88(2)  & 0.511(2)  & 1.39(2)  \\
640 & 0.40740 &   4.0021(18)    & 3.0347(4)  & 2.521(2)  &  1.07(3)  & 0.514(2)  & 1.45(3)  \\
800 & 0.40210 &   3.9952(19)    & 3.2221(3)  & 2.704(2)  &  1.20(3)  & 0.518(2)  & 1.50(3)  \\
\hline
\end{tabular}
\end{center}
\caption{Results for the first few cumulants and their differences for fixed physical volume $m(L)L = 4$. The bare parameters
$\delta$ are taken from \cite{arXiv:1009.2146}.}
\label{tab:parameters}
\end{table}

\begin{figure}
\begin{center}
\begin{tabular}{cc}
\hspace{-1cm} \includegraphics[width=8cm]{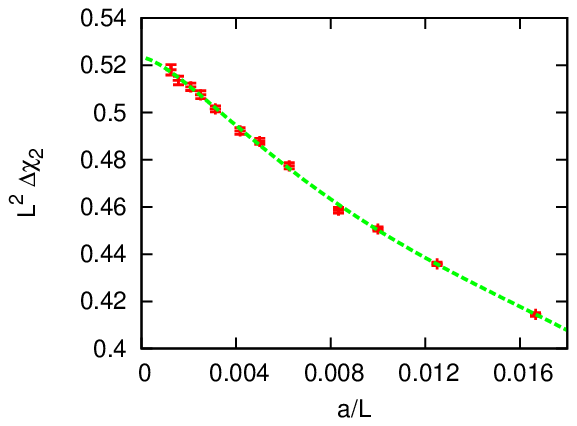} 
\hspace{-1cm} \includegraphics[width=8cm]{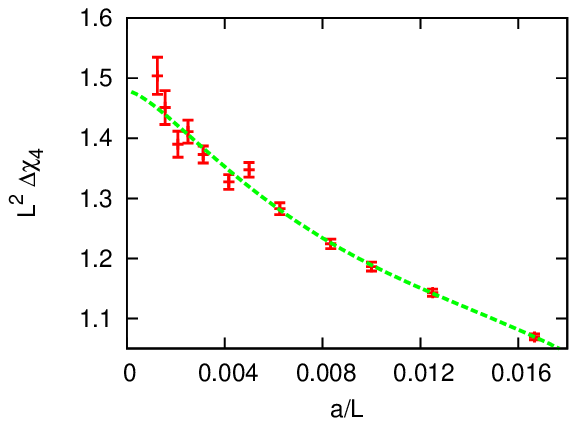} 
\end{tabular}
\caption{Continuum extrapolation for the first two cumulant differences multiplied by the volume, $L^2 \Delta \chi_2$ and $L^2 \Delta \chi_4$.}
\label{fig:cumulantdiffs}
\end{center}
\end{figure}

The results for the cumulant differences $\Delta \chi_2$ and $\Delta \chi_4$ are shown on figure \ref{fig:cumulantdiffs}.
Obtaining continuum estimates is not entirely trivial since the precise form of the leading and sub leading cut-off effects is not
known a priori. Using the results of \cite{Balog:2009np, Balog:2009yj} one may expect the leading corrections to be $O((a/L)^2)$
with possibly large logarithmic corrections. Fits of the form
\bea
\label{fits}
C + (a/L)^2 \left( \sum_{j=n}^m A_j \log^j(L/a) \right)
\eea
with $(n,m)=(0,3), (1,3), (2,3), (0,2)$ all work quite well with $\chi^2/{\rm dof}$ values close to unity for $\Delta\chi_2$
and slightly higher, around $1.8$ for $\Delta\chi_4$. The continuum extrapolated values agree in both cases among the four fit
function choices and the four curves lie almost entirely on top of each other. In both cases the $(n,m)=(0,2)$ choice is shown 
on the plots leading to continuum estimates $C=0.523(2)$ and $1.48(2)$ for $L^2 \Delta \chi_2$ and $L^2 \Delta \chi_4$, respectively.
Clearly, both values are non-zero.

\section{Summary and conclusion}
\label{summary}

It has been known for a long time that the topological susceptibility in the two dimensional $O(3)$ model is ill-defined in the
continuum. Consequently the topological charge distribution $P(Q)$ does not have a finite continuum limit. The semi-classical
analysis predicts precisely what part of $P(Q)$ is actually divergent and what part of it is finite. In this work
non-perturbative evidence has been presented supporting the semi-classical result. The only divergent
quantity is the first Fourier coefficient of the free energy density,
\bea
f_1 = - \int_0^\pi f(\theta)\cos(\theta) \frac{d\theta}{2\pi}\;,
\eea
while the remaining part $\sum_{n>1} (1-\cos(n\theta)) f_n$ is finite and non-zero. Hence the quantity 
\bea
f_R(\theta) = f(\theta) - ( 1 - \cos(\theta) ) f_1
\eea
is finite and universal and one may consider the subtraction an additive renormalization. Similarly the renormalized partition function 
$Z_R(\theta) = \exp( - V f_R(\theta) )$ is finite and universal and related to the bare partition function by a multiplicative
renormalization. Instead of subtracting $f_1$ it is sufficient to subtract only its divergent piece.
The logarithmic singularity is expected to be volume independent\footnote{I thank Ferenc Niedermayer for pointing this
out.}. Let us then denote this singular quantity by $f_{1s}$. Since $\chi_2 = f_1 + \sum_{n>1} n^2 f_n$ a suitable definition of
$f_{1s}$ is the logarithmic singularity in the topological susceptibility which can directly be measured in lattice calculations. 
A natural renormalization procedure is then the following: one defines the theory for non-zero $\theta$ by the action
\bea
S(\theta) = S(\theta = 0) - i\theta Q - (1-\cos(\theta))Vf_{1s}
\eea
and all resulting correlation functions related to topology (i.e. derivatives with respect to $\theta$) become finite. The last term in the
full action above is a non-perturbatively generated counter term.
It is important to note that the above renormalization does not mean that $\theta$ itself gets renormalized, the
bare $\theta$ is still a physical quantity which does not require renormalization. 
It would of course be very interesting to check the volume independence of $f_{1s}$ in lattice simulations.

The finite quantities $f_{n>1}$ and $\Delta\chi_{2m}$ are not volume independent and are non-trivial functions of $z = m(L)L$. 
Since the model is exactly solvable at $\theta =
0$ it would be interesting to derive the first few cumulant differences $\Delta\chi_{2m}(z)$ from the exact solution or at least
their value in the infinite volume limit.

In any case the finite and non-zero cumulant differences naturally lead to the conclusion that $\theta$ is a relevant coupling of the theory and the total
topological charge operator $Q$ is a relevant operator despite the ill-defined nature of the moments $\langle Q^{2m}\rangle$.

The original motivation was the study of a toy model mimicking confining, walking and conformal behavior in four dimensional gauge
theories in order to study the severe systematic effects of the latter. It was proposed that increasing $\theta$ is analogous 
to increasing the number of flavors $N_f$ because as
$\theta$ goes from zero to $\pi$ the model goes from confining to walking and to conformal. In the toy model a suitable
renormalized coupling is $g_R^2(L) = m(L)L$ which would then run with the finite volume $L$. A necessary condition 
for this analogy to hold was establishing precisely the divergence structure of the partition function at non-zero $\theta$.

A particular difficulty of the gauge theory calculation can also be studied in the toy model. It is very difficult to distinguish
numerically the following two cases: the theory with zero quark mass just below the conformal window and the theory with a small
but non-zero quark mass just inside the conformal window. Both theories walk, the former for the usual reason of
being just below the conformal window while the latter because even though it would be conformal for zero quark mass, the
non-zero mass drives the coupling away from the would-be fixed point as soon as the running scale goes below the massive fermionic states.
This phenomenon can be mimicked in the toy model by considering it at zero external magnetic field and $\theta = \pi -
\varepsilon$ and also at a small but non-zero external magnetic field and $\theta = \pi$. Both theories are expected to walk and it
would be interesting to explore in the toy model what intrinsic features are different despite the similar behavior of the walking
coupling constant.

There are a couple of differences between the toy model and gauge theory though. Less important is the fact that while $\theta$
does not enter the perturbative $\beta$-function, $N_f$ does. More significant is the fact that due to the $\theta \to -\theta$
symmetry and periodicity by $2\pi$ the two values $\theta - \varepsilon$ and $\theta + \varepsilon$ lead to the same continuum
theory and it does not have an infra red fixed point (for non-zero $\varepsilon$ the coupling walks). This means that the zero of the
$\beta$-function at $\theta=\pi$ is eliminated by arbitrary perturbations of $\theta$ meaning that this zero is a second order
zero, unlike in the gauge theory where generically the zero is expected to be first order and is
preserved by small perturbations. Hence the $\theta=\pi$ model is really
analogous to a gauge theory which is exactly at the lower edge of the conformal window.
It would be interesting to find a simple toy model which possesses all essential features and in addition the infra red fixed
point is a first order zero of the $\beta$-function and disappears by joining with a non-trivial UV fixed point as expected in
gauge theory \cite{Gies:2005as, Kaplan:2009kr, Braun:2010qs}.

\section*{Acknowledgments}

I am grateful for very helpful discussions with J\'anos Balog, \'Arp\'ad Heged\H us, S\'andor Katz, Julius Kuti, Martin Luscher and Ferenc Niedermayer.
I am especially grateful to Ferenc Niedermayer for sharing his cluster algorithm source code.

This work is supported by the EU Framework Programme 7 grant (FP7/2007-2013)/ERC No 208740 and in part
by the National Science Foundation under Grant No. NSF PHY05-51164. I thank the KITP Santa Barbara for hospitality while preparing
the manuscript.

\end{document}